\documentstyle[prb,aps,eqsecnum,multicol]{revtex}
\newcommand{\beq}{\begin{equation}}
\newcommand{\eeq}{\end{equation}}
\newcommand{\beqy}{\begin{eqnarray}}
\newcommand{\eeqy}{\end{eqnarray}}
\newcommand{\bm}[1]{{\mbox{\boldmath ${#1}$}}}

\newcommand{\bx}{{\bf x}}
\newcommand{\bp}{{\bf p}}
\newcommand{\bu}{{\bf u}}
\newcommand{\bR}{{\bf R}}
\newcommand{\hbx}{{\hat{\bx}}}
\newcommand{\hbp}{{\hat{\bp}}}
\newcommand{\Ham}{{\hat{H}}}
\newcommand{\bk}{{\bf k}}
\newcommand{\bq}{{\bf q}}
\newcommand{\bA}{{\bf A}}
\newcommand{\e}{{\cal E}}
\newcommand{\D}[2]{\frac{d {#1}}{d {#2}}}
\newcommand{\PD}[2]{\frac{\partial {#1}}{\partial {#2}}}

\newcommand{\SS}[1]{\scriptscriptstyle #1}
\newcommand{\dotprod}{\stackrel{\SS{\bullet}}{{}}}
\newcommand{\pr}[3]{Phys.\ Rev.\ {\bf #1}, #2 (19#3)}
\renewcommand{\prb}[3]{Phys.\ Rev.\ B {\bf #1}, #2 (19#3)}
\renewcommand{\prl}[3]{Phys.\ Rev.\ Lett.\ {\bf #1}, #2 (19#3)}
\renewcommand{\rmp}[3]{Rev.\ Mod.\ Phys.\ {\bf #1}, #2 (19#3)}
 \newcommand{\PRSL}[3]{Proc.\ Royal.\ Soc.\ (London) {\bf #1}, #2 (19#3)}
\begin{document}
\draft
\title{Wave-packet dynamics in slowly perturbed crystals:
               Gradient corrections \\ and Berry-phase effects}
\author{Ganesh Sundaram and Qian Niu}
\address{Department of Physics,
University of Texas at Austin \\ Austin, Texas 78712-1081}
\date{\today}
\maketitle
%\tighten
\begin{abstract}
We present a unified theory for wave-packet dynamics of electrons
in crystals subject to perturbations varying slowly in space and
time. We derive the wave-packet energy up to the first order
gradient correction and obtain all kinds of Berry-phase terms for
the semiclassical dynamics and the quantization rule. For
electromagnetic perturbations, we recover the orbital
magnetization energy and the anomalous velocity purely within a
single-band picture without invoking inter-band couplings. For
deformations in crystals, besides a deformation potential, we
obtain a Berry-phase term in the  Lagrangian due to lattice
tracking, which gives rise to new  terms in the expressions for
the wave-packet velocity and the  semiclassical force. For
multiple-valued displacement fields surrounding dislocations,
this term manifests as a Berry phase, which we show to be
proportional to the Burgers vector around each dislocation.
\end{abstract}
\begin{multicols}{2}
%\narrowtext
%\baselineskip=12pt
\section{Introduction}
\label{Intro}
%%%%%%%%%%%%%%%%%%%%%%%%%%%%%%%%%%%%%%%%%%%
Our understanding of electronic properties of crystalline solids
is primarily based on the Bloch theory  for periodic
systems.\cite{ashcroft} It has been of great interest to extend
this theory to situations where crystals  are perturbed in various
ways. So far the most useful description has been the
semiclassical theory for electron dynamics within a band
supplemented by the semiclassical quantization rule or the
Boltzmann transport equations. For example, the equations of
motion of Bloch electrons in electromagnetic fields are given
by\cite{ashcroft-12}
\beqy
\dot{\bx} &=&  \ \ \,
\frac{1}{\hbar}\PD{\e_{0,n}(\bk)}{{\bf k}}, \nonumber \\ \hbar
\dot{\bk} & = & - e\, {\bf E}- e\, \dot{\bx} \times {\bf B},
\label{eom-em1}
\eeqy
where $\e_{0,n}(\bk)$ is the energy of the
$n$th band of an unperturbed crystal. These equations have played
a fundamental role in the  physics of metals and semiconductors.

The derivation of Eq.~(\ref{eom-em1}) dates back to Bloch, Peierls,
Jones and Zener in the early 1930's.\cite{wpd-J34} By assuming
that the transition probabilities to other bands are negligible, they
showed that the Eqs.~(\ref{eom-em1}) describe the motion of a
narrow wave packet obtained by superposing the Bloch states
of a band. Various extensions of the theory have been made to deal
with  perturbations of more general nature and to obtain
corrections to Eqs.~(\ref{eom-em1}) in high fields.

Peierls\cite{Peierls-H} pioneered the effort of constructing
an effective one-band Hamiltonian to describe the
quantum dynamics of a Bloch electron. By using the
tight-binding model, he was able to  show that the effective
Hamiltonian in the presence of a magnetic field may be obtained by
replacing the crystal momentum $\hbar\bk$ by the gauge invariant
momentum operator  $[-i \hbar \nabla  + e \bA(\hbx)]$ in the
unperturbed band energy:
\beq
\label{peierls-sub}
 \Ham_{\rm eff}
= \e_{0,n}\left[-i \nabla + \frac{e}{\hbar} \bA(\hbx)\right],
\eeq
which later came to be known as the Peierls substitution.
Two decades later, Slater\cite{Slater-H} and
Luttinger\cite{Luttinger-H} gave a more rigorous derivation of the
effective Hamiltonian for electromagnetic perturbations, by
expanding  the wave function in the basis of Wannier functions:
\beq
\Psi(\bx,t) = \sum_{l} {f}_{l}(t) \: W(\bx-\bR_l),
\eeq
where $\{\bR_l\}$ are the lattice positions. They showed that the
envelope function  ${f}(\bx,t)$, defined by
${f}(\bR_l,t) = {f}_{l}(t)$ and a smooth interpolation between the
atomic positions, satisfies the effective  Schr\"{o}dinger equation
\beq
\label{eff-S-eqn}
i \hbar \PD{}{t} f=
\left\{\e_{0,n}\left[-i\nabla + \frac{e}{\hbar}\bA(\bx) \right] - e
\phi(\bx) \right\} f,
\eeq
where $\phi(\bx)$ is a slowly varying scalar potential. The
equations of motion (\ref{eom-em1}) then follow from
Eq.~(\ref{eff-S-eqn})  and the correspondence principle.

 Further development of the theory was made by taking into account
the effects of inter-band coupling.
Adams\cite{Adam-mb}  extended  the works of Slater and
Luttinger  to many band operator formalism.  Karplus,
Luttinger, and Kohn  derived a correction to the velocity, known as
the anomalous velocity,  and predicted  a spontaneous Hall effect  in
ferromagnetic materials.\cite{KLK-av}  Later, Adams and
Blount\cite{AB-59,Blount} showed that this term
arises from the noncommutability between the Cartesian
components  of the intra-band position
operator. Recently, Chang and Niu\cite{Chang1,Chang2} related the
anomalous velocity correction to the Berry phase associated
with the electron motion in an energy band.\cite{Berry,Simon,Zak}
Corrections to the effective Hamiltonian as an asymptotic series in
the  field strength were obtained by eliminating the inter-band
matrix  elements with unitary transformations by
Kohn, Blount, and Roth in the early 1960's.\cite{eff-ham}
Later, Brown\cite{Brown-67} extended the Wannier function method
to  crystals under high magnetic fields  using magnetic translation
symmetry.
A decade later, a variant of the Wannier function method that treats
position and momentum in a symmetric way, known as
$kq$-representation, was developed by Zak.\cite{Z-kq-rep}
Recently, Rammal and Bellissard used an algebraic
approach\cite{Bellissard} and Wilkinson an operator
approach\cite{Wil-Heff}
to derive the first order field correction for the special case of
the Harper's equation.\cite{Harper,incom-S85}

Another approach to this problem involves the use of the
WKB expansion to derive a Hamilton-Jacobi equation,
and then making the correspondence from the classical variables to
the quantum operators.
This method was applied by many researchers to understand the
nature of the spectrum and the wave function of electrons described
by the Harper's equation, particularly in the semiclassical
limit.\cite{incom-S85}  A more general treatment of the problem is
based on a two-scale  expansion in which the electron coordinate
and the slowly varying  vector potential are regarded as independent
variables.\cite{Nenciu}

 In this work we come back to the original consideration of a
wave packet in a band and  use a time dependent variational
principle\cite{tdvp-K,argendu} to derive the wave-packet energy up
to  first order in the gradient of the perturbations and
Berry-phase corrections to the semiclassical dynamics and
the quantization rule. We are able to obtain the magnetization
energy and the anomalous  velocity entirely from the single-band
point of view. Also, our method can be directly extended to the case
of  slowly perturbed magnetic bands, where methods based on
the usual Wannier function approach break down because of
nonexistence  of localized Wannier functions for such bands in
general.\cite{Gen-Wf}

 This program was started with Chang and Niu\cite{Chang1,Chang2}
for the  special case of a two-dimensional periodic system in a
strong magnetic  field. Here we establish a unified framework for
slowly perturbed  crystals whose Hamiltonian can be expressed in
the form
\beq
\label{I.H}
H[\hbx, \hbp; \beta_1(\hbx,t),\ldots, \beta_r(\hbx,t)],
\eeq
where $\{ \beta_i(\bx,t) \}$ are the modulation functions
characterizing the perturbations. They may represent either
deformation strain fields, gauge potentials of electromagnetic
fields, or slowly varying impurity potentials. They also appear
in model potentials for modulated and incommensurate
crystals,\cite{Balki} and for graded semiconductors.\cite{Geller}

We shall illustrate our formalism with two special cases of
perturbations:  electromagnetic fields and deformations in
crystals. In the first case, in addition to the corrections of
orbital magnetization energy and anomalous velocity mentioned
above, we discuss the Peierls substitution, Berry-phase-modified
Landau levels, and Zak-phase-modified Wannier-Stark levels. For
deformational  perturbations, we show that the deformation
correction to the wave-packet energy can be obtained from the
differential shift in the band energy under uniform strain.
Then we obtain for the Lagrangian a Berry-phase term due to
lattice tracking, which gives rise to
new terms in the expressions for the wave-packet velocity
and the semiclassical force; for multiple-valued displacement
fields in the presence of dislocations, this term manifests as a
Berry phase, which we show to be proportional to the Burgers vector
around each dislocation, and thus in a sense keeps track of the
lattice position. We also discuss the consequences of the Berry-phase
term on electron transport, and
the Aharonov-Bohm type effects in dislocated crystals.

The paper is organized as follows. We present our formalism in
Sec.~\ref{Formalism},  treat electromagnetic  and deformational
perturbations in  Secs.~\ref{EMF} and \ref{DEFM} respectively, and
conclude  with a summary in Sec.~\ref{Summary}.

\section{Formalism}
\label{Formalism}
We shall begin by constructing a basis local to the wave packet and
describe the wave packet in detail. Then we derive the Lagrangian,
the semiclassical equations  of motion, and Berry-phase correction
to the semiclassical quantization, and discuss some aspects of
formal quantization through the Hamiltonian formalism.
\subsection{The local basis and the wave packet}
Consider a wave packet centered at $\bx_c$ at a given time, with
its spread small compared to the length scale of the perturbations.
Then the approximate Hamiltonian that the wave packet ``feels''
may be obtained by linearizing the perturbations about the wave
packet center as
\beq
\label{Happrox}
\Ham \approx \Ham_c +
\left\{ \begin{array}{c}
\mbox{sum of terms }\propto \\
(\hbx - \bx_c) \cdot \mbox{grad}_{\bx_c} \, \beta_i(\bx_c,t)
\end{array}\right\},
\eeq
where $\Ham_c \equiv H[\hbx,\hbp; \{\beta_i(\bx_c,t)\}]$ will be
called the local Hamiltonian. The terms within the braces are small
in the  neighborhood of the wave packet and may be treated
perturbatively. The local Hamiltonian $\Ham_c$ has the required
periodicity of the unperturbed crystal, and has an energy
spectrum of bands (Bloch bands) with Bloch wave eigenstates
satisfying
\begin{equation}
\label{eval_eq}
 \Ham_c(\bx_c,t) |\psi_\bq({\bx_c,t}) \rangle =
{\cal E}_c(\bx_c,\bq,t) |\psi_\bq({\bx_c,t})\rangle,
\end{equation}
where $\bq$ is the Bloch wave vector and ${\cal E}_c(\bx_c,\bq,t)$
is the band energy. Since we will be concerned with only a single
band, we have omitted the band index for simplicity of notation.
We note that both the wave-packet center $\bx_c$ and time $t$
enters  in the Bloch states and the band energy parametrically. We
shall see that the dependency on the center position of the wave
packet will manifest as new types of Berry-phase terms in the
equations of  motion.

These eigenstates form a convenient basis to expand the wave packet.
Specifically, we write
\begin{equation}
\label{wp}
\left| \Psi \right>=\int \! d^3\!q \:
a(\bq,t) |\psi_\bq({\bx_c,t})\rangle,
\end{equation}
where $a(\bq,t)$ is the amplitude with the normalization
\beq
\label{norm-q}
\int d^3\!q \, |a(\bq,t)|^2  = \langle \Psi | \Psi \rangle = 1.
\eeq
Here we have used the convention that
\beq
\langle \psi_{\bq}| \psi_{\bq'}\rangle = \delta(\bq-\bq').
\eeq
 It is assumed that the distribution $|a(\bq,t)|^2$ is narrow compared
to the size of the Brillouin zone and has the mean wave vector
\begin{equation}
\label{wpk}
 \bq_c  =   \int \! d^3\!q \, \bq\, |a(\bq,t)|^2.
\end{equation}
To be consistent, the wave packet must yield the pre-assigned center
position:
\begin{equation}
\label{wpx}
\bx_c  =  \left<\Psi\right| \hbx \left| \Psi \right>.
\end{equation}
This condition can be expressed in terms of  other wave-packet
parameters as follows. Writing the amplitude in the
form $a(\bq,t) = |a(\bq,t)| \,  \exp [-i\gamma(\bq,t)]$, and
using the matrix elements of position operator $\hbx$ between
the Bloch states [Eq.~(\ref{x-m_e-n}) of Appendix  \ref{A-Ham}],
we find that
\beq
\label{x-exp}
\left<   \Psi\right| \hbx  \left| \Psi \right>
=  \int d^3q \, |a|^{2}
\left[\PD{\gamma}{\bq} + \left< u\biggl| \biggr. i
\PD{u}{\bq} \right> \right],
\eeq
where $ |u\rangle \equiv |u(\bx_c,\bq,t)\rangle =
e^{-i\bq\cdot \hbx} |\psi_\bq (\bx_c,t) \rangle $ is the periodic
part of the Bloch wave, and inner products involving the periodic
part $| u \rangle$ mean an integration over the unit-cell volume
$v_c$ with a factor of $(2\pi)^3/v_c$, which implies the normalization
$\langle u | u \rangle = 1 $. According to our
assumption of a narrow  wave packet in the
$\bq$-space, Eq.~(\ref{wpx}) becomes
\beq
\label{pos}
\bx_c= \PD{\gamma_c}{\bq_c} +
\left< u\biggl| \biggr. i \PD{u}{\bq_c} \right>,
\eeq
where $|u\rangle$ now stands for $|u(\bx_c,\bq_c,t)\rangle$, and
$\gamma_c$  for $ \gamma(\bq_c,t)$.

In writing down the expansion (\ref{wp}), we have assumed that the
wave packet that is initially in a band always lies in the same
band. This is justified if the band is separated from other bands by
finite gaps, and if the time and the length scales of the
perturbations are long compared to those associated with these
gaps.\cite{condition}

\subsection{The lagrangian and dynamics}
The dynamics of the mean position $\bx_c$ and the crystal
momentum $\hbar\bq_c$ can in principle be derived from the
Schr\"{o}dinger equation for the wave packet. It is more
conveniently obtained using a  time dependent variational
principle\cite{tdvp-K,argendu} with the Lagrangian given by
\beq
\label{effL}
L = \left<\Psi \right|i\D{}{t}-\Ham \left|\Psi \right>,
\eeq
where and hereafter we use the convention $\hbar =1$.
We use ${d}/{dt}$ to mean the derivative with respect to
the time dependence of the wave function explicitly or
implicitly through $\bx_c$ and $\bq_c$. The partial
derivative ${\partial }/{\partial t}$, is reserved for those with
$\bx_c$ and $\bq_c$ held fixed.

Under the previously discussed conditions on the widths of the wave
packet, we can evaluate the Lagrangian as a function of $\bx_c$ and
$\bq_c$,  their time derivatives,  and the time $t$:
\beq
\label{Lag-app}
L \approx L(\bx_c, \dot{\bx}_c, \bq_c, \dot{\bq}_c,t).
\eeq
The terms involving higher moments of the wave packet, which
specify its width and shape, are of higher order  in the
gradient of the perturbations and hence are neglected.\cite{WPD}

Accordingly, we obtain for the first term
in Eq.~(\ref{effL}),
\beq
\label{timederiv}
\left< \Psi \left|i \D{\Psi}{ t}\right. \right>
   =    \frac{\partial \gamma_c }{\partial t}  +
 \left< u \biggl| \biggr. i\PD{ u}{t}\right>
+  \dot {\bx}_{c} \cdot \left<u\biggl| \biggr. i\PD{u}{\bx_{c}}
\right>.
\nonumber
\eeq
The first term comes from the explicit time
dependence of $\gamma_c$.
The contribution from $|a(\bq,t)|$ is zero because of the
normalization condition (\ref{norm-q}) on the amplitude.  The
second and third  terms come  about because
of the dependence of the  basis functions  on time explicitly and
implicitly through
$\bx_c$ respectively. Using the relation
\beq
\label{tot-deriv}
\PD{\gamma_c}{t} = \D{\gamma_c }{t}
  - \dot {\bq}_{c}\cdot  \PD{\gamma_c}{\bq_{c}},
\eeq
and Eq.~(\ref{pos}), Eq.~(\ref{timederiv}) can be recast into the
form
\beqy
\lefteqn{\left< \Psi \left|i \D{\Psi}{ t}\right. \right> =
\frac{d\gamma_c }{dt} -\dot{\bq}_{c} \cdot \bx_c
      +  \dot{\bq}_{c} \cdot \left< u\biggl|
      \biggr. i\frac{\partial u}
      {\partial \bq_{c}}  \right>}   \hspace{2.5cm}
\nonumber  \\
   & &  +\dot {\bx}_{c} \cdot\left< u\biggl| \biggr.
i\frac{\partial u} {\partial \bx_{c}} \right>
+\left< u\biggl| \biggr. i\PD{u }{t} \right>,
\eeqy
where $\gamma_c$ appears only in a total time derivative.

The expectation value of the Hamiltonian, which gives the
wave-packet energy $\e$, may be evaluated up to first order in
the perturbation gradients using the
linearized Hamiltonian (\ref{Happrox}):
\beqy
\e   =  \langle \Psi | \Ham | \Psi \rangle
      \approx  \langle \Psi | \Ham_c | \Psi \rangle +
      \langle \Psi | \Delta \hat{H} | \Psi \rangle,
\eeqy
where  the gradient correction $\Delta \hat{H}$
 may be written as\cite{symmetrzn}
\beq
\label{g-c-op}
\Delta \hat{H}  = \frac{1}{2} \left[ (\hbx - \bx_c) \cdot
\PD{\Ham_c}{\bx_c} + \PD{\Ham_c}{\bx_c}\cdot(\hbx -
\bx_c)\right].
\eeq
The expectation value of the local
Hamiltonian is just the  band energy at the mean wave vector,
\beq
\langle \Psi | \Ham_c | \Psi \rangle = \e_c(\bx_c,\bq_c,t),
\eeq
while the gradient correction requires some
calculations  [Appendix \ref{A-Ham}], but the result has the
simple form
\begin{equation}
\label{grad-corr}
\Delta {\cal E} =  - \mbox{Im}\left. \left[
\left<\frac{\partial u}{\partial \bx_c}\right| \dotprod  ({\cal E}_c -
 \Ham_c) \left|\frac{\partial u}{\partial \bq} \right>
\right]\right|_{\bq = \bq_c},
\end{equation}
where ``$\dotprod$'' denotes scalar product between the
vectors formed by gradients with respect to $\bx_c$ and $\bq$.

 The Lagrangian thus takes the form
\beqy
\label{Lgen}
 L = & - & \e   +   {\bq}_{c} \cdot  \dot\bx_c
      +  \dot{\bq}_{c}\cdot\left< u\biggl| \biggr. i\PD{u}{\bq_{c}}
\right>
\nonumber \\
 & +  &  \dot {\bx}_{c} \cdot \left< u\biggl| \biggr. i\PD{u }{
\bx_{c}}\right> +\left< u\biggl| \biggr. i\PD{u}{t} \right>.
\eeqy
where we have neglected a term of total time derivative
${d(\gamma -\bx_c \cdot \bq_c)}/{dt}$ in the  Lagrangian, as it
does not affect the equations of motion and the quantization rule.
The last three terms may be grouped into a single
term $\langle u | i \, d u / dt \rangle$, which turns out to be the net
rate of change of Berry phase for wave-packet motion within the
band. We note that under the transformations $|u\rangle
\rightarrow \exp[i \varphi(\bx_c,\bq,t)] \, |u\rangle $ or
$\bq \rightarrow \bq + {\bf K}$,
${\bf K}$ being a reciprocal lattice vector, the Lagrangian remains
invariant up to a total time derivative of some
function of $\bx_c$, $\bq_c$, and $t$. The former corresponds to
gauge invariance while the latter to periodicity in the reciprocal
space.

From the Lagrangian (\ref{Lgen}) we obtain the following
equations of semiclassical motion:
\beqy
\dot{\bx}_{c}  & = & \ \ \, \PD{\e}{ \bq_c} -  \left(
\tensor{\Omega}_{\bq\bx}
\cdot \dot{\bx}_c  +  \tensor{\Omega}_{\bq\bq} \cdot \dot{\bq}_c
\right)  + \bm{\Omega}_{t\bq},
\nonumber  \\
\dot{\bq}_c  & = & -\PD{\e}{\bx_c}  +
\left( \tensor{\Omega}_{\bx\bx} \cdot \dot{\bx}_c +
\tensor{\Omega}_{\bx\bq}\cdot \dot{\bq}_c \right) -
\bm{\Omega}_{t\bx}. \label{gsceom}
\eeqy
The components of the tensor $\tensor{\Omega}_{\bq\bq}$ are defined
by
\beq
\label{BCkk}
 (\tensor{\Omega}_{\bq\bq})_{\alpha \beta} \equiv
\Omega_{{q_\alpha}{q_\beta}} \equiv i
\left[ \left< \frac{\partial u}{\partial q_{c\alpha}}\biggl| \biggr.
 \frac{\partial u}{\partial q_{c \beta}} \right>-
\left< \frac{\partial u}{\partial q_{c \beta}}\biggl| \biggr.
        \frac{\partial u}{\partial q_{c\alpha}} \right> \right],
\eeq
and those of the vector $\bm{\Omega}_{t\bx}$ by
\beq
\label{BCtx}
 (\bm{\Omega}_{t\bx})_\alpha \equiv {\Omega}_{tx_\alpha} \equiv
 i \left[ \left< \frac{\partial u}{\partial t}\biggl| \biggr.
        \frac{\partial u}{\partial x_{c\alpha}} \right>-
 \left< \frac{\partial u}{\partial x_{c\alpha}}\biggl| \biggr.
        \frac{\partial u}{\partial t} \right> \right],
\eeq
where $\alpha$ and $\beta$ are Cartesian indices.  The other tensors
$ \tensor{\Omega}_{\bx \bx}, \,
\tensor{\Omega}_{\bx \bq}$, and $\tensor{\Omega}_{\bq \bx}$ and  the
vector $\bm{\Omega}_{t\bq}$ are defined similarly. These quantities are
known as Berry curvatures.\cite{Berry,Simon} We note that these
equations involve Berry curvatures between every pair of parameters
and that they  have  symplectic symmetry in the absence of time
dependence.
\subsection{Formal and semiclassical quantization}
We mentioned in the Introduction that the equations of motion  were
usually derived from the effective Hamiltonian upon using the
correspondence principle. Here we consider the reverse process to
obtain the effective quantum Hamiltonian from the
semiclassical dynamics. This requires a knowledge of
canonical structure of the wave-packet dynamics. Following the
standard procedure of analytical mechanics, we introduce the
canonical momenta conjugate to the generalized coordinates
\beqy
{\bf P}_1 &=& \PD{L}{{\dot \bx}_c} = \bq_c +
\left< u\biggl| \biggr. i\PD{u}{ \bx_{c}} \right>, \label{cmom1} \\
{\bf P}_2 & = &  \PD{L}{\dot\bq_c} =
\left< u\biggl| \biggr. i\PD{u}{\bq_c} \right>, \label{cmom2}
\eeqy
and the semiclassical Hamiltonian ${\cal H}$ by the Legendre
transformation
\beqy
{\cal H} & = & \dot\bx_c \cdot {\bf P}_1 + \dot\bq_c \cdot {\bf P}_2 - L
                             \nonumber \\
                 & = &  \e(\bx_c,\bq_c,t) -
\left< u\biggl| \biggr. i \PD{u}{t} \right> \label{Hsc}.
\eeqy
The semiclassical Hamiltonian is independent of ${\bf P}_1$ and
${\bf P}_2$, because  the Lagrangian  is linear in the generalized
velocities.

Starting with ${\cal H}$, regarded formally as a function of
$\bx_c$, $\bq_c$, ${\bf P}_1$, and ${\bf P}_2$, one cannot obtain
the equations of motion (\ref{gsceom}) from the Hamilton equations.
This is because Eqs.~(\ref{cmom1}) and (\ref{cmom2}) defining
${\bf P}_1$ and ${\bf P}_2$ do not depend on the generalized velocities,
and hence they should be treated as constraints between the canonical
variables.

In the simple case where the Berry-phase terms are zero, these
constraints become ${\bf P}_1=\bq_c$, and ${\bf P}_2=0$, and ${\cal H}
= \e$.  These suggest that we treat $(\bx_c,\bq_c)$ as a canonical
pair and forget about the other  degrees of freedom. By doing so, we can
indeed obtain the equations of motion (\ref{gsceom}) from the Hamilton
equations. Having identified the canonical pair, one can proceed with a
formal procedure of quantization (``requantization"), $\bq_c
\rightarrow -i\partial / \partial \bx_c$, to obtain an effective quantum
Hamiltonian. A slightly more general case will be encountered in
the case of electromagnetic  perturbations in the next section of
this article.

When Berry-phase terms are present, constraints (\ref{cmom1}) and
(\ref{cmom2}) still imply some hidden canonical relations between
$\bx_c$ and $\bq_c$,  but these are entangled in a complicated manner
that cannot be expressed explicitly in general. This clearly shows the
difficulty of the Hamiltonian formalism in the presence of Berry-phase
terms. If one insists on using the Hamiltonian approach, one can employ
the method of  Lagrange multipliers, which allows the  spurious degrees
of freedom to be formally treated as  independent,\cite{Dirac} and
obtain the equations of motion (\ref{gsceom}).  The ``requantization"
procedure for this case is  quite complicated and needs further
investigation.

The semiclassical quantization, on the other hand, is quite straight
forward. In order that stationary states and energy levels can be
talked about, we shall restrict ourselves to static perturbations. For
a wave-packet motion that is  regular and is described by closed
orbits in the phase space $(\bx_c,\bq_c)$, semiclassical energy
levels are obtained using  the  quantization procedure\cite{tabor}
due to Einstein, Brillouin, and  Keller,
\beq
\label{EBK}
\oint_{{\cal C}} {\bf P}_1\cdot d\bx_c + \oint_{{\cal C}}
  {\bf P}_2\cdot d\bq_c = 2\pi [m + \frac{\nu}{4}],
\eeq
where  ${\cal C}$ denotes an orbit of constant energy $\e$, $m$ an
integer that labels the eigenvalue, and $\nu$ the number of caustics
traversed. With Eqs.~(\ref{cmom1}) and (\ref{cmom2}) for
${\bf P}_1$ and ${\bf P}_2$, the above condition  reduces to
\beq
\oint_{{\cal C}} \bq_c \cdot d\bx_c = 2\pi \left[ m +
\frac{\nu}{4}-\frac{\Gamma({\cal C})}{2\pi} \right],
\eeq
where
\beq
\label{Bphase3}
\Gamma({\cal C})
 = \oint_{{\cal C}} d\bx_{c} \cdot
\left< u\biggl|\biggr. i\frac{\partial  u }
{\partial \bx_{c}} \right>  +  \oint_{{\cal C}} d\bq_c \cdot
     \left< u\biggl| \biggr. i\frac{\partial  u } {\partial
 \bq_{c}}   \right>
\eeq
is the Berry phase acquired by the wave packet upon going round the
closed orbit once. The Berry-phase correction to the  quantization
condition  made its first appearance  in Wilkinson's
work\cite{Wil-Bp} on Harper's equation, in Kuratsuji and Iida's
work\cite{pathIQR} on adiabatic nuclear motion, and also recently
in the work of Chang and Niu\cite{Chang2} on wave-packet dynamics
in magnetic Bloch bands.

\section{Electromagnetic fields}
\label{EMF}
So far our treatment of perturbations has been in general terms,
and our results are in an abstract form. Their physical meaning will
become clear through the consideration of two special cases in this
and the next section. For a class of perturbations for which the
Hamiltonian is of the special form
\beq
\label{sp-form}
H_0[\hbx+\bm{\beta}_1(\hbx,t), \hbp + \bm{\beta}_2(\hbx,t)] +
\beta_3(\hbx,t),
\eeq
all the results can be expressed in terms of the unperturbed Bloch
wave  basis. In this section we shall consider electromagnetic
perturbations for which $\bm{\beta}_1(\hbx,t) = 0$.
\subsection{The gauge invariant crystal momentum}
Let $\Ham_0(\bq)$ denote the Hamiltonian for the bare crystal,
with the eigenstate $|u(\bq)\rangle$ (the periodic part of the Bloch
wave) and the  band  energy ${\cal E}_0(\bq)$ for a particular band.
The Hamiltonian gets modified by the gauge potentials $[\bA(\bx,t),
\phi(\bx,t)]$ of an electromagnetic field to
\beq
\label{ham-em}
\Ham = H_0[\bq+e\bA(\hbx,t)]  -e\phi(\hbx,t).
\eeq
This has  the form (\ref{I.H}) and (\ref{sp-form}),
 with the gauge potentials playing the role
of the modulation functions, and hence the local Hamiltonian
must have the form
\beq
\Ham_c = \Ham_0[\bq+e\bA(\bx_c,t)]  -e\phi(\bx_c,t).
\eeq
As $e\bA(\bx_c,t)$ is only an additive constant to the
crystal momentum $\bq$, the  basis states have the form
$|u(\bx_c, \bq, t) \rangle = |u(\bk) \rangle $, where $\bk = \bq + e
\bA(\bx_c,t)$ is the gauge invariant or mechanical crystal
momentum. In terms of the gauge  invariant crystal momentum
$\bk$, a number of simplifications can be obtained. First, the
eigenenergy can be written in the form
\beq
\label{em-energy}
{\cal E}_c(\bx_c,\bk,t) ={\cal E}_0(\bk) - e\phi(\bx_c,t).
\eeq
Second, the gradient correction (\ref{grad-corr})
becomes the orbital magnetization energy of the wave packet,
\beq
  - {\bf M} \cdot {\bf B},
\eeq  where ${\bf B} \equiv {\rm curl}_{\bx_c} \bA(\bx_c,t)$
is the magnetic field, and
\beq
\label{mag-mom}
{\bf M}  =  -e \, {\rm{Im}} \left. \left[
\left<\PD{u}{\bk} \right|
\times ({\cal E}_0-\Ham_0(\bk))
\left| \PD{u}{\bk} \right> \right]\right|_{\bk = \bk_c}
\eeq
is the orbital magnetic moment of  Bloch electrons. Third, the last
three terms of the Lagrangian (\ref{Lgen}) simply become the
single term $\dot{\bk}_{c}\cdot\langle u | i \partial u/ \partial
{\bk_c} \rangle$.  Finally, the Lagrangian takes the form
\beqy
\label{em-lag}
L= -  \e_{\SS{M}} &+& e\phi(\bx_c,t) + \dot{\bx}_{c}\cdot \bk_c
  \nonumber \\
&-  &    e\, \dot{\bx}_{c}\cdot  \bA(\bx_c,t)
+\dot{\bk}_{c}\cdot\left< u \biggl| \biggr. i\PD{u}{\bk_{c}} \right>,
\eeqy
where $\e_{\SS{M}}  \equiv {\cal E}_0(\bk_c)  -
{\bf M}\cdot {\bf B}$.
\subsection{The reciprocal magnetic field and orbital magnetization
energy}
The equations of motion can either be derived variationally from the
above Lagrangian or directly from Eqs.~(\ref{gsceom})
derived for the general case in the previous section. They have the
form
\beqy
\dot{\bx}_{c}  & = &  \ \ \, \PD{\e_{\SS{M}}}{{\bf k}_c}
-   \dot{\bk}_c \times \bm{\Omega},   \nonumber \\
\dot{\bk}_c   &=&   -  e\, {\bf E} - e\, \dot{\bx}_c \times {\bf B},
\label{eomef}
\eeqy
where
\( {\bf E} \equiv - {\rm grad}_{\bx_c} \phi(\bx_c,t)  -
\partial {\bA(\bx_c,t)}/ \partial {t} \)
is the electric field, and
\beq
\label{rec-mag}
(\bm{\Omega})_\alpha \equiv
\frac{1}{2} \epsilon_{\alpha \beta \gamma}
 (\tensor{\Omega}_{\bk \bk})_{\beta \gamma},
\eeq
are the components of the vector form of the antisymmetric tensor
$\tensor{\Omega}_{\bk\bk}$ given by (\ref{BCkk}).  In
(\ref{rec-mag}) and henceforth, repeated Cartesian indices are
taken to be summed.  Because
$\bm{\Omega}$  occupies a similar position as the magnetic field
in the equations of motion, it will be called the reciprocal
magnetic field.

 Equations of motion (\ref{eomef}) differ from Eqs.~(\ref{eom-em1}) in two
respects. Firstly, the energy $\e$ contains a correction term
from the magnetic moment (\ref{mag-mom}) of Bloch electron.
This term has been derived earlier as a first order correction in the
theory of Bloch electrons subject to  magnetic
fields.\cite{mag.en}  A similar  term has also been found in the
theory of electrons in incommensurate
lattices,\cite{Bellissard,Wil-Heff}  and  in the theory of wave
packet dynamics in magnetic Bloch bands.\cite{Chang2} Secondly,
the correction term to the velocity,
$-\dot{\bk}_c\times\bm{\Omega}$, is the anomalous velocity
that was predicted to give rise to a spontaneous Hall conductivity in
ferromagnetic materials.\cite{KLK-av,AB-59}
In the context of the quantum Hall effect, the integral of the Berry
curvature ${\Omega}_{k_\alpha k_\beta}$ over the Brillouin zone
was shown to be proportional to the Hall conductivity for a full band
and to be quantized (Chern's topological invariant).\cite{tktn}
Recently, Chang and Niu\cite{Chang2} proved this result
semiclassically. It seems more appropriate to call this term  Hall
velocity than anomalous velocity.

 The semiclassical equations (\ref{eomef}) should be
invariant under time reversal, spatial inversion, or
certain rotations if these are symmetries of the unperturbed
crystal. Such symmetries impose severe restrictions on the
behavior of the  reciprocal magnetic field $\bm{\Omega}$ and the
magnetic moment ${\bf M}$ as functions of $\bk$. Under time
reversal, $\dot{\bx}_c$, $\bk_c$, and ${\bf B}$ change sign while
$\bx_c$, $\dot{\bk}_c$, and ${\bf E}$ are invariant. If the bare
crystal is  invariant under time reversal, we must have
$\bm{\Omega}(-\bk) = -\bm{\Omega}(\bk)$, and  ${\bf M}(-\bk) =
-{\bf M}(\bk)$,   which implies in particular that they must vanish
at $\bk = 0$.  Under spatial inversion, ${\bf E}$, $\bx_c$, $\bk_c$,
and the time derivatives of the last two change sign
while ${\bf B}$ remains unchanged. If the bare crystal
has inversion symmetry, we must have
$\bm{\Omega}(-\bk) = \bm{\Omega}(\bk)$, and
${\bf M}(-\bk) = {\bf M}(\bk)$. Finally, if the system is
invariant under certain proper rotations,  the reciprocal
magnetic field and the  magnetic moment should transform
like vectors under these rotations.

For monatomic non-magnetic crystals, both time reversal and
spatial  inversion symmetries are present, rendering $\bm{\Omega}$
and ${\bf M}$ null everywhere in the Brillouin zone. However, it is
not entirely justified to ignore these quantities for  magnetic
crystals or non-magnetic crystals without inversion symmetry
(such as GaAs). Investigations have been undertaken to see whether
the presence of the reciprocal magnetic field and the orbital
magnetization lead to observable effects.

 The Lagrangian (\ref{em-lag}) and the equations of motion
(\ref{eomef}) were derived earlier by Chang and Niu\cite{Chang2}
for perturbed magnetic Bloch electrons in two dimensions in the
gauge where $\phi(\bx,t) = 0$. They used a wave packet that
gauges away the vector potential locally so that their magnetic
Bloch wave vector is the same as the corresponding gauge
invariant crystal momentum $\bq+e\bA(\bx_c,t) = \bk$ here. Their
derivation was less general in that it is only for two dimensions
and  more general in that it provides a description of electrons in
rational magnetic fields ${\bf B}_0$, for which the flux through a
unit cell equals a rational fraction [$\sim {\cal O}(1)$] of the flux
quantum
$(h/e)$. However, our formalism can easily be generalized to this
situation, for a more general three dimensional case, by assuming a
background of constant rational magnetic field ${\bf B}_0$. All we
need to do is to interpret $\bq$ as the wave vector of magnetic Bloch
states defined within a reduced Brillouin zone (the magnetic
Brillouin zone),\cite{Chang2} and to replace ${\bf B}$ in
Eq.~(\ref{eomef}) by ${\bf B} - {\bf B}_0$.

\subsection{Peierls substitution and Landau levels}
It follows  from the Lagrangian (\ref{em-lag}) that the canonical
momenta are given by
\beqy
{\bf P}_1 & = & \bk_c - e\bA(\bx_c,t), \label{cm-em1}\\
{\bf P}_2  & = & \left<  u \biggl| \biggr. i\PD{ u}{\bk_c} \right>,
\label{cm-em2}
\eeqy
and the Hamiltonian by  \( {\cal H} = \e_{\SS{M}}(\bk_c) -
e \phi(\bx_c,t) \).
In the absence of the Berry-phase term (\ref{cm-em2}),  we may
obtain the
Hamiltonian as a function of the canonical pair
$(\bx_c, {\bf P}_1)$ as
\beq
{\cal H} = \e_{\SS{M}}[{\bf P}_1 + e\bA(\bx_c,t)] -e \phi(\bx_c,t).
\eeq
 The
quantization of this Hamiltonian by setting
${\bf P}_1 = -i \partial / \partial \bx_c$ amounts to the
Peierls substitution. However, it is not clear how to deal
with the case with a Berry-phase term using the Hamiltonian
approach.

When only  a uniform magnetic field  is present, the two equations
of motion (\ref{eomef}) can be combined into a single one for the
$\bk$-space motion. It is evident that a $\bk$-space orbit must
lie in a plane normal to ${\bf B}$ and must be on a constant energy
surface of $\e(\bk)$. If such an orbit is closed, known as a
cyclotron orbit, the EBK formula yields
\beq
\frac{1}{2} \hat{{\bf B}}\cdot \oint_{\cal C} \bk_c \times d\bk_c =
\frac{e |{\bf B}|}{\hbar}\left( m+ \frac{1}{2} - \frac{\Gamma({\cal
C})}{2\pi} \right),
\label{lanlev}
\eeq
where
\beq
\label{Bp-k}
\Gamma({\cal C}) = \oint_{\cal C} d\bk \cdot \left< u\biggl| \biggr.
i\PD{u}{
\bk_{c}}  \right>
\eeq
is the Berry phase accumulated by the wave packet upon completing
a circuit along the loop ${\cal C}$, and we have restored the Planck
constant. The left-hand side of Eq.~(\ref{lanlev}) is just the
$\bk$-space area enclosed by the orbit
${\cal C}$. As this phase influences energy levels, it affects the
density of states. It is shown in Ref.~\onlinecite{Chang2} that
$\Gamma$ plays an important role in determining the spectral
splitting pattern of magnetic bands.
\subsection{Zak phase and  Wannier-Stark ladder}
In this subsection, we restrict our discussion to one dimension for
simplicity. The semiclassical motion  under a uniform electric
field is described by the Hamiltonian ${\cal H} = \e_{0}(k_c)+
eEx_c$. It follows from  the boundedness and  the periodicity of
the band energy  that the motion in real space is also  bounded and
periodic.  Such a closed  motion in the real space is known as
Bloch oscillations.  In the reduced  zone scheme, the motion is
closed also in the phase space
$(x_c,k_c)$, which is a cylinder. Quantizing this motion according
to Eq.~(\ref{EBK}) gives the  condition
\beq
\label{wslqrule}
- \int\limits_{-{\pi}/{a}}^{{\pi}/{a}} \! \! \!\! d k_c \, x_c(k_c)   =
2\pi \left( m + \frac{\nu}{4} - \frac{\Gamma}{2\pi} \right),
\eeq
where $a$ stands for the lattice constant, and
\beq
\label{zakph}
\Gamma= \int\limits_{-{\pi}/{a}}^{{\pi}/{a}} \!\!\! \! dk
\, \left< u  \biggl| \biggr. i\frac{\partial u}{\partial k}\right>
\eeq
is known as the Zak phase,\cite{zak-ph} and $x_c(k_c)$ is the
constant energy curve for the $m$th energy level defined by
$W_m = \e_0(k_c) + eEx_c$,  $m$ being an integer. Averaging this
expression over the orbit, we obtain from Eq.~(\ref{wslqrule})
\beq
\label{wslad}
W_m =  \bar{\e}_{0} + e E a \left(-m  -\frac{\nu}{4} +
\frac{\Gamma}{2\pi} \right),
\eeq
where  $\bar{\e}_{0}$ is the average of the band energy over the
Brillouin zone, and $m$ is any integer between
$-\infty$ to $\infty$, since the mean value of $x_c$ can be
anywhere on the cylinder. This spectrum, known as the
Wannier-Stark ladder, was first derived by Wannier\cite{WSL}
without the Berry-phase term. The correction was due to Zak, who
later interpreted it as a  Berry phase.\cite{zak-wsl-Bp}
\section{Deformations in Crystals}
\label{DEFM}
We shall now come to deformational perturbations. It turns out that
the model Hamiltonian for a deformed  crystal  also has the special
form (\ref{sp-form}) with $\bm{\beta}_2(\hbx,t) = 0$, which is
$H_0[\hbx+\bm{\beta}_1(\hbx,t), \hbp] + \beta_3(\hbx,t)$,
and hence all the corrections are expressible in terms of
the undeformed basis for this case too.

\subsection{The translated crystal basis}
 A deformed crystal with atomic displacements
$\{ \bu_l \}$ may be described by the Hamiltonian\cite{lax,abs-dp}
\beq
\label{H-defm}
H = \frac{\hbp^2}{2m} + V_0[\hbx-\bu(\hbx)] + s_{\alpha
\beta}(\hbx) {\cal V}_{\alpha \beta}[\hbx - \bu(\hbx)],
\eeq
where $\bu(\bx)$ is a smooth displacement field\cite{df-el}
satisfying $\bu(\bR_l + \bu_l) = \bu_l$, and
$s_{\alpha \beta} = \partial u_\alpha / \partial x_\beta$
is the unsymmetrized strain.
 The justification of the above Hamiltonian and an explicit
expression for the last term are given in Appendix
\ref{A-d_pot}. While the last term of the Hamiltonian, being
proportional to the strain, can be treated perturbatively, the
other terms are of the form of Eq.~(\ref{I.H}), with the displacement
field playing  the role of the modulation function. The local
Hamiltonian is then given by
\beq
\label{ham-loc-d}
\Ham_c = \frac{\hbp^2}{2m} + V_0[\hbx - {\bf u}(\bx_c)].
\eeq
This is nothing but the Hamiltonian of an undeformed crystal
shifted in position by the displacement field at the center of the
wave packet, $\bu(\bx_c)$.  The band energy is therefore the same
as that of the undeformed crystal, ${\cal E}_0(\bk)$, and the
eigenstates are the translated Bloch waves
$\{ \psi_\bk[\bx - \bu(\bx_c)] \}$.

 Our wave packet will thus be formed out of these translated Bloch
states of the undeformed but translated crystal. This procedure is
valid so long as the strain is weak, so that the variation of the
displacement within the spatial width of the wave packet is small.
When the first order corrections to the Hamiltonian are taken into
account, our method should give the same physical results (to the
same order) as obtained using  a strained basis. However, our
formulation should be simpler  and easier to interpret, because it
avoids the necessity of transformation between the lab and lattice
frames of reference.

Although the small strain regime covers the vast majority of
practical situations, it is some times necessary to consider the
effect  of large strains. In the case where a large uniform and
static strain is  superposed on top  of a small varying strain, our
formulation  can still be applied; one  only needs  to interpret the
basis as that of the uniformly strained crystal. There can be a third
possibility in which the strain variation  is large  over large
distances but is small over the size of the wave packet. In this
case,  it is more appropriate to use a strained local basis,  that is, a
basis of a homogeneously strained crystal with the   strain value
given by the actual strain at  the center of the  wave packet.

\subsection{The crystal deformation potential}
The  wave-packet energy is obtained by summing the expectation
values of the local Hamiltonian (\ref{ham-loc-d}), the gradient
correction (\ref{g-c-op}), and the last term of Eq.~(\ref{H-defm}).
Because of the functional form of the basis states,  we have
\beq
\langle \Psi | \Ham_c |\Psi \rangle = \e_0(\bk_c),
\eeq
while the gradient correction becomes
\beq
\label{gc-defm}
 - m s_{\alpha \beta}(\bx_c)
[ \langle\hat{v}_\alpha \hat{v}_\beta \rangle  -
\langle  \hat{v}_\alpha \rangle
\langle \hat{v}_\beta \rangle ].
\eeq
The angular  brackets in the above expression represent  the
expectation value of the enclosed operators in the
Bloch state at $\bk=\bk_c$, and
$\hat{v}_\alpha  = \partial \Ham(\bk) / \partial k_\alpha$ is the
velocity operator.  As is well known,
$\langle \hat{v}_\alpha \rangle = \partial \e_0  /
\partial k_\alpha \equiv v_\alpha$ is the group velocity of
Bloch electrons.
As for the last term of the Hamiltonian (\ref{H-defm}), we may
write, in accordance with our approximation
[Eq.~(\ref{Lag-app})],
\beq
\label{strain-corr}
\langle \Psi | s_{\alpha \beta}(\hbx)
\hat{\cal V}_{\alpha \beta} | \Psi \rangle
\approx s_{\alpha \beta}(\bx_c)
\langle \hat{\cal V}^c_{\alpha \beta} \rangle ,
\eeq
where ${\cal V}^c_{\alpha \beta} =
{\cal V}_{\alpha \beta}[\bx - \bu(\bx_c)]$.
We again write the energy of the wave packet in the form
\beq
\label{wpe-d}
\e  = \e_0(\bk_c) + \Delta \e(\bx_c,\bk_c),
\eeq
where, this time, the correction $\Delta \e$, which is to
be called the deformation potential for the wave
packet,\cite{D-P} has two contributions
[Eqs.~(\ref{gc-defm}) and (\ref{strain-corr})]. As the latter is
proportional to the local deformation, it may be written
in the form
\beq
\label{defm-pot}
\Delta \e =  s_{\alpha \beta}\, (\bx_c)
D^{w}_{\alpha \beta}(\bk_c),
\eeq
with
\beq
\label{dpc-w}
D^{w}_{\alpha \beta} = m [  v_\alpha v_\beta -
\langle\hat{v}_\alpha \hat{v}_\beta \rangle ] +
\langle \hat{\cal V}^c_{\alpha \beta}\rangle.
\eeq
We note that this quantity vanishes in the free electron limit
($V_0 \rightarrow 0$),
which is quite satisfying from a physical point of view:
a wave packet should not feel the effect of a
deformation of the lattice in  the absence of
electron-lattice coupling.

Further, from the differential band structure\cite{DPT-KA}
under a uniform strain $\tensor{\epsilon}$,
\beqy
{\cal E}'(\bk')-{\cal E}_0(\bk)
& = & \epsilon_{\alpha  \beta} D^b_{\alpha \beta}(\bk) \nonumber \\
& = & \epsilon_{\alpha \beta}[- m\langle\hat{v}_\alpha
\hat{v}_\beta \rangle
+ \langle \hat{\cal V}^c_{\alpha \beta}\rangle],
\label{bs_dpc_us}
\eeqy
where
$\bk' = (\tensor\openone + {\tensor{\bf s}})^{-1} \cdot \bk$,
 $\Delta \e$ can be determined by the relation
\beq
\label{dpt-wb}
D^{w}_{\alpha \beta} = D^b_{\alpha \beta} + m v_\alpha v_\beta,
\eeq
according to Eqs.~(\ref{dpc-w}) and (\ref{bs_dpc_us}).
The same expression has been
obtained for the deformation potential for electron-phonon
interaction at long wavelengths, showing the
equivalence of the latter with our wave-packet
deformation potential (\ref{defm-pot}).\cite{D-P-T}

\subsection{Equations of motion and lattice tracking}
Before we proceed further, let us first make the displacements
time  dependent, $\{ \bu_l(t) \}$, as we are discussing the
dynamical aspect.
Accordingly, we extend the results of the previous subsections
with the replacements $\bu(\bx) \rightarrow \bu(\bx,t)$ and
${s}_{\alpha \beta}(\bx) \rightarrow {s}_{\alpha \beta}(\bx,t)$,
and with the Bloch wave basis given by the states
$\{ \psi_\bk(\bx - \bu(\bx_c,t))\}$.

As for the Lagrangian, the functional form of the basis states
imply that the last two terms of its general expression
(\ref{Lgen}) become
\beq
\label{d-g-pot}
\left<   u \biggl| \biggr. i \PD{u}{\bx_c} \right>  =
{f_\alpha}\PD{u_\alpha}{\bx_c}, \quad
%\nonumber \\
\left<   u\biggl| \biggr.  i \PD{u}{t}\right>  =
{ f_\alpha} \PD{{u}_\alpha}{t},
\eeq
where
\beq
\label{Bp_fact}
{\bf f}(\bk) = \frac{m}{\hbar} \PD{\e_0}{\bk}  - \hbar \, \bk,
\eeq
in which $\hbar$ has been restored. The reader is warned that
the displacement field ($\bu$ or $u_\alpha$) should not be
confused with the periodic part of the Bloch state, $|u\rangle$.
The quantity ${\bf f}(\bk)$, being the difference between the group
momentum of a Bloch electron and the momentum of a free electron,
denotes that part of the momentum arising from
the lattice interaction alone. It has the desired property
of vanishing in the free electron limit, where the lattice
deformation should not be felt. By substituting Eq.~(\ref{d-g-pot}) into
Eq.~(\ref{Lgen}), we obtain
\beq
\label{lag-defm}
L = - \e + \dot\bx_c \cdot \bk_c + \dot\bk_c \cdot \left<  u\biggl|
\biggr. i \PD{u}{\bk_c}\right>  + \dot{\bf u} \cdot {\bf f}(\bk_c),
\eeq
where $\dot\bu$ denotes  $d \bu / dt$. The last term is new and
represents the rate of change of the Berry phase due to lattice
tracking.  We shall see below that there is a tendency for the lattice
to drag the electron with its displacement motion, hence the word
``tracking''.  This term also gives rise to the Burgers vector when
integrated around a dislocation, so in a sense it keeps track of the
lattice position.

Similarly, the Berry curvatures in Eqs.~(\ref{gsceom}) take values
\beqy
{\Omega}_{k_\alpha x_\beta}  &=&  -{\Omega}_{x_\beta
k_\alpha}   =   {\PD{u_\gamma}{x_{c\beta}}
\PD{{f}_\gamma}{k_{c\alpha}}}, \nonumber \\
 {\Omega}_{t k_\alpha} &=&
{- \PD{{f}_\gamma}{k_{c\alpha}} \PD{u_\gamma}{t},} \quad
 {\Omega}_{x_\alpha x_\beta}    =    {\Omega}_{t x_\beta}  =  0,
\label{defm-omegas}
\eeqy
and by expressions (\ref{wpe-d})
and (\ref{defm-pot}) for $\e$, the equations of motion become
\beqy
{\dot x}_{c\alpha} & = & \PD{\e_0}{k_{c\alpha}}   +
s_{\beta \gamma} \PD{D^{w}_{\beta \gamma}}{k_{c\alpha}}  -
 (\dot\bk_c \times \bm{\Omega}  )_\alpha   -
\dot{u}_\beta \PD{f_\beta}{k_{c\alpha}},
 \\ {\dot k}_{c\alpha} &=& \quad \quad \: \, -
D^{w}_{\beta \gamma} \PD{s_{\beta \gamma}}{x_{c\alpha}} \:  -
s_{\beta \alpha}
\PD{f_\beta}{k_{c\gamma}} {\dot  k}_{c\gamma}.
\label{eom-defm}
\eeqy

In the above equations, the semiclassical force has two
contributions: the first term arises from the deformation
potential, and the second term, which arises from the lattice
tracking term, will be called the tracking force.
For uniform strains the semiclassical force vanishes,
so that $\bk_c$ is a good quantum number in such a case.

The velocity has three contributions in addition to the usual term
given by the gradient of the band energy. The second term
corresponds to the group velocity arising from the deformation
potential, and the third term, as was seen earlier in
Eq.~(\ref{eomef}) of section \ref{EMF}, is the Hall velocity.
The last term arises from the lattice tracking term of the
Lagrangian, and will be similarly called the tracking velocity.
To understand this term better, let us rewrite the term
using Eq.~(\ref{Bp_fact}) as
\beq
\label{v-adia}
\dot{u}_\alpha - m \dot{u}_\beta  \frac{\partial^2 \e_0}{\partial
k_{c\beta}
\partial k_{c\alpha}}.
\eeq
In the free electron limit, where electrons and lattice
decouple, the adiabatic velocity vanishes identically, for
\beq
\frac{\partial^2 \e_0}{\partial k_{c\alpha} \partial  k_{c\beta}}
\rightarrow
\frac{\delta_{\alpha \beta}}{m}.
\eeq
On the other hand, when the band under consideration is full, the
second term of (\ref{v-adia}) averaged over the band is zero,
implying a complete adiabatic following of the lattice motion, which
confirms an earlier result on adiabatic particle transport.\cite{niu-ct}
\subsection{Dislocations and Berry phase}
Our formalism is also applicable to dislocation strain fields, which
are well defined except in a region of a few atomic spacings around
the line of dislocation. Outside this region, the displacement field
$\bu(\bx)$ is a smooth but multiple-valued function. The change in
the displacement field along a closed loop around the line of
dislocation,
\(
\label{Burgers-v}
\Delta \bu = \oint_{\cal C} dx_\alpha \partial \bu / \partial
x_\alpha = {\bf b},
\)
is known as the Burgers vector; it equals one of the Bravais
lattice vectors. On account of
this multiple-valuedness, a wave packet of incident wave vector
$\bk$ taken around the line of  dislocation  acquires a Berry phase
\beq
\label{Bp-disl}
\Gamma = \oint_{\cal C} d\bx_c\! \cdot \left< u \biggl| \biggr.
i\PD{u}{\bx_c} \right> = \oint_{\cal C} \! d \bu\cdot{\bf f}(\bk_c)
\approx {\bf b}\cdot{\bf f}(\bk),
\eeq
where we have assumed $\langle u | i\, \partial u / \partial \bk
\rangle$ to be zero\cite{bc-zero} and the  corrections arising from
the changes in the wave vector to be negligible.
In other words the action integral of the last
term of (\ref{lag-defm}) around a dislocation gives us a
Berry phase that is proportional to the Burgers vector.

Note that this Berry phase is independent of the path as long as it
encloses the dislocation line.
What we have is a  situation similar to the
Aharonov-Bohm effect,\cite{A-B-effect} with the dislocation
playing the role of the solenoid, and the Berry curvature
$\bm{\Omega}_{\bx\bx}$ the role of the magnetic field. Just as in
the case of the solenoidal field, $\bm{\Omega}_{\bx\bx} = 0$
everywhere except for the core region where it should be taken to
be singular.

The Berry phase (\ref{Bp-disl}) affects the scattering
of electrons by a dislocation,\cite{A-B-scatt} and our result can be
used to compute the shift in the scattering fringe pattern due the
Berry phase. The intensity distribution has interference terms
of the form $\cos(\theta + \Gamma)$ due to each pair of paths along
the two sides of the dislocation, where $\theta$ is given by the
path length difference per wavelength of the incident beam.
The Berry phase and hence the shift are maximal when the incident
wave vector $\bk$ is such that ${\bf f}(\bk) = m {\bf v} - \hbar \bk$
is parallel to the Burgers vector ${\bf b}$.  To fix ideas, let us
make the assumption that ${\bf f} \, \| \, \bk$. For an edge
dislocation, where ${\bf b}$ lies in the normal plane of
the dislocation axis, we expect that maximal effect of the Berry
phase is seen when the direction of the incident beam is
perpendicular to the axis and coincide with the direction of
${\bf b}$. For a screw dislocation, the maximal Berry phase occurs
when $\bk$ is parallel to the axis along which ${\bf b}$ lies.
However, since the beam must pass the dislocation in order to
produce the interference pattern, the maximal Berry phase effect
should actually occur when $\bk$ is along some finite angle away
from the axis.

    The Berry phase can also affect the electron diffraction pattern
of a deformed crystal with or without a dislocation.  When an
electron beam of wave vector $\bk$ is sent through a crystal, it
propagates as Bloch waves in different bands of the crystal, all
with the same Bloch wave vector which is equal to $\bk$ modulo a
reciprocal lattice vector.  Because of the energy differences of the
bands, the Bloch waves grow out of phase from one another as they
propagate.  Therefore, when they exit the crystal, they recombine to
produce not only the incident beam but also the diffracted beams.
For a deformed crystal, the phase change of a Bloch wave is given by
the time integral of the Lagrangian (\ref{lag-defm}).  The
contribution from the $\dot \bx_c \cdot \bk_c$ term may be
dropped, because it is independent of the band index.  For weak
strains, we may also neglect the deformation potential and the third
term in Eq.~(\ref{lag-defm}).  The dominant correction to the dynamical
phase arises from the last term: $\Delta \bu \cdot {\bf f}(\bk)$,
where $\Delta \bu$ is the change in the displacement field over the
path of propagation of the Bloch waves.  To show the
soundness of these general ideas, we have calculated the
diffraction pattern from a thin slab of crystal containing a screw
dislocation, correctly reproducing  earlier
experimental and theoretical results.\cite{Bp-Sdisl}
\section{Summary}
\label{Summary}
We provide a unified framework for wave-packet
dynamics of electrons in slowly perturbed crystals, which consists of
constructing a wave packet using the Bloch states belonging to a
single band  of the local periodic Hamiltonian, obtained by
replacing the perturbations by their value at the center of
the wave packet, and deriving its dynamics in a general form,
based on the time dependent variational principle.
We derive the wave-packet energy up to
first order in the gradient of the perturbations and all kinds of
Berry-phase corrections to the semiclassical dynamics and the
quantization rule. Also, we give a discussion of a formal
quantization procedure through the Hamiltonian formalism
with the semiclassical
dynamics as the starting  point.

We illustrate our framework with two cases of perturbations.
For electromagnetic fields, previous results  of orbital
magnetization and anomalous velocity are obtained purely from a
single-band point of view.  For deformations in crystals,
we obtain a Berry-phase term in the  Lagrangian due to lattice
tracking, which gives to the new terms of tracking velocity and
force in the equations of motion of the wave packet. For
multiple-valued displacement fields in the presence of
dislocations, this term manifests as a Berry phase, which we show
to be proportional to the Burgers vector around each dislocation.
Also, we relate the deformation correction to the wave
packet energy to the shift in band energy under uniform
strain, which turns out to be the same as the
deformation potential for electron-phonon
interaction at long wavelengths.

       The combined effects of electromagnetic and deformational
perturbations are yet to be studied.  Given that the perturbed
Hamiltonian is of the form (\ref{sp-form}), the equations of motion
for the wave-packet dynamics can be completely determined in
terms of the properties of the unperturbed crystal and the fields of
perturbations. Such a theory will provide a basis for electron
transport in deformed crystals and should be pursued in the near
future.

\acknowledgements
The authors wish to  thank M.~C.~Chang, R.~Diener, E.~Demircan,
S.~Fishman, D.~Gavenda, G.~A.~Georgakis, L.~Kleinman, M.~Marder,
F.~Nori, R.~Resta,  E.~C.~G.~Sudarshan, Bala Sundaram, and J.~Zak
for many insightful  discussions.  This work was supported by NSF
(Grant Nos.~DMR 9705406 and PHY  9722610), and by the
R.~A.~Welch Foundation.
\appendix
%%%%%%%%%%%%%%%%%%%%%%%%%%%%%%%%%%%%%%%%%%%%%%%%%
\section{\protect{\\}The gradient correction $\Delta \e$}
\label{A-Ham}
In the derivation below, the band indices for the
Bloch states and the energy bands have been restored.

 We begin with the matrix elements of the position
operator,\cite{Blount}
\beq
\label{x-m_e-n}
 \langle \psi_{n\bq}|\hbx| \psi_{n'\bq'}\rangle  =
\left[\, i \, \PD{}{\bq} \delta_{nn'} +  \left< u_n \left| i
\PD{u_{n'}}{\bq} \right. \right>
\, \right] \, \delta(\bq' - \bq),
\eeq
and  the identity
\beqy
 \label{h_g-m_e}
\lefteqn{\left< \psi_{n\bq} \left| \PD{\Ham_c}{\bx_c}
\right| \psi_{n'\bq'} \right> = }   \nonumber \\
 & & \hspace{.3cm}  \delta(\bq - \bq') \left<  u_n \left|
\PD{\Ham_c}{\bx_c}
\right|  u_{n'} \right>  = \\
& &  \hspace{.6cm}
\delta (\bq -\bq')\left[ ( \e_{cn} - \e_{cn'} )
\langle \PD{u_{n\bq}}{\bx_c} | u_{n'\bq} \rangle +
\PD{\e_{cn}}{\bx_c} \delta_{nn'} \right], \nonumber
\eeqy
which are easily verified.

 From Eqs.~(\ref{x-m_e-n}), (\ref{h_g-m_e}), and the completeness
relation
\beq
\sum_n \int d^3q \: |\psi_{n\bq} \rangle \,
\langle \psi_{n\bq} | = \hat{\openone},
\eeq
 we can  show that
\beqy
\label{h_g-x-me}
\lefteqn{ \left< \psi_{n\bq} \left| \PD{\hat{H}_c}{\bx_c}\cdot
\hbx \right| \psi_{n\bq'}\right> =
\PD{\e_{cn}}{\bx_c}\cdot \langle \psi_{n\bq}|\hbx|\psi_{n\bq'}\rangle +
  } \nonumber \\
& & \hspace{1.3cm} i \left< \PD{u_{n\bq}}{\bx_c}\biggl|
\dotprod (\e_{cn} -
\Ham_c)  \biggr|  \PD{u_{n\bq}}{\bq} \right> \delta(\bq-\bq').
\eeqy

If we assume that the wave packet is constructed from the
Bloch states of $n$th band,  we have
\beq
\label{exp-der}
\left< \Psi \left| \PD{\Ham_c}{\bx_c}\right| \Psi \right>
= \int d^3 q \, |a|^2  \, \PD{\e_{cn}}{\bx_c} =
\left.\PD{\e_{cn}}{\bx_c}\right|_{\bq = \bq_c},
\eeq
as only the second term of Eq.~(\ref{h_g-m_e}) contribute to the
expectation value.
By the same token, from Eqs.~(\ref{exp-der}) and (\ref{h_g-x-me}),
we obtain
\beqy
\label{exp-h_g-x}
\lefteqn{\left< \Psi \left| \PD{\Ham_c}{\bx_c} \cdot (\hbx - \bx_c)
 \right| \Psi \right> \hspace{.6cm} =} \hspace{.5cm} \\
&& \hspace{1.2cm}
i \left. \left< \PD{u_{n\bq}}{\bx_c}\biggl| \dotprod (\e_{cn} -
\Ham_c)  \biggr|  \PD{u_{n\bq}}{\bq} \right> \right|_{\bq = \bq_c}.
\nonumber
\eeqy
Hence the expectation value (\ref{grad-corr}) of
Eq.~(\ref{g-c-op}) is just half of the sum of (\ref{exp-h_g-x})
and its complex conjugate.
%\narrowtext
\section{\protect{\\} The Deformed Crystal Potential}
\label{A-d_pot}
Given the atomic displacements $\{\bu_l\}$, which could be due to
either strain or rotations, let us denote the deformed crystal
potential regarded as a function of the coordinates of the electron
and the equilibrium atomic positions by $V(\bx; \{\bR_l + \bu_l\})$.
If the displacements are small, we can expand the potential in
powers of $\{\bu_l\}$ and do away with the perturbation theory.
When they are large, we have to adopt a different technique. For this
purpose we introduce a smooth displacement field $\bu(\bx)$ such
that $\bu(\bR_l+\bu_l) = \bu_l$, and write the potential as
\beq
V[\bx - {\bf u}(\bx); \{\bR_l+ {\bf u}_l - {\bf u}(\bx)\}],
\eeq
where we have used the invariance property of a potential under a
simultaneous translation of electronic and atomic positions. We may
now  expand the potential in powers of ${\bf u}_l - {\bf u}(\bx)$ as
\beq
\label{defmcpot}
V_0[\bx - {\bf u}(\bx)] + \sum_l
[{\bf u}_l - {\bf u}(\bx)]_\alpha {V}_{l \alpha} + \cdots.
\eeq
where $V_0(\bx - \bu) \equiv V[\bx-\bu;\{\bR_l\}]$ is the potential
used in the deformable ion model in which the approximated
potential at $\bx$ equals the undeformed crystal potential at the
undeformed electron coordinate $\bx - {\bf u}(\bx)$, and
\beq
{V}_{l\alpha} \equiv
\PD{V[\bx - \bu(\bx); \{\bR_l \}]}{R_{l\alpha}}
 = {V}_{l\alpha} [\bx - \bu(\bx) - \bR_l].
\eeq
The equality in the last equation is meant to suggest that the center
of fall off of ${V}_{l\alpha}$ is at the zero of the expression
$\bx - \bu(\bx) - \bR_l$, viz., $\bx = \bu_l + \bR_l$.

The expansion (\ref{defmcpot}) is meaningful only if ${V}_{l\alpha}$
decreases sufficiently rapidly with increasing $|\bx - \bu(\bx) -
\bR_l|$. This condition holds for metals because of Coulomb
screening; it has been argued that ${V}_{l\alpha}$ is short
ranged\cite{KSP-Resta} also for non-polar semiconductors and
insulators.

Under such a condition, we
may write
\beqy
u_{l\alpha} &=&  u_\alpha (\bx + \bR_l +
\bu_l - \bx) \nonumber \\
 &\approx& u_\alpha (\bx) + (\bR_l +
\bu_l - \bx)_\beta s_{\alpha \beta}(\bx) \nonumber \\
&\approx& u_\alpha(\bx) +
[\bR_l + \bu(\bx) - \bx]_\beta s_{\alpha \beta}(\bx),\label{bu-exp}
\eeqy
whence the summation in Eq.~(\ref{defmcpot}) can be put into the form
$s_{\alpha \beta}{\cal V}_{\alpha \beta}$ with
\beq
\label{pert-coef}
{\cal V}_{\alpha \beta}[\bx-\bu(\bx)]
 = \sum_{l} [\bR_l + \bu(\bx)- \bx]_\beta {V}_{l\alpha}.
\eeq
%\vspace{-.5cm}

\end{multicols}

\begin{references}
\bibitem{ashcroft}
N.~W.~Ashcroft and N.~D.~Mermin, {\it Solid State Physics}
(W.~B.~Saunders Co., Philadelphia, 1976) Chaps.~8-11.
\bibitem{ashcroft-12}
See  {\it Solid State Physics}, (Ref.~\onlinecite{ashcroft}), Chap.~12.
\bibitem{wpd-J34}
H.~Jones and C.~Zener, \PRSL{144}{101}{34}.
References to works of Bloch and Peierls, and their
approximations have been given here.
\bibitem{Peierls-H}
R.~Peierls, Z.~Physik {\bf 80}, 763 (1933).
\bibitem{Slater-H}
J.~C.~Slater, \pr{76}{1592}{49}.
\bibitem{Luttinger-H}
J.~M.~Luttinger, Phys.~Rev.~{\bf 84}, 814 (1951).
\bibitem{Adam-mb}
E.~N.~Adams, \pr{85}{41}{52}.
\bibitem{KLK-av}
R.~Karplus and J.~M.~Luttinger, \pr{95}{1154}{54};
W.~Kohn and J.~M.~Luttinger, {\it ibid} {\bf 108}, 590 (1957);
J.~M.~Luttinger, {\it ibid} {\bf 112}, 739 (1958).
\bibitem{AB-59}
E.~N.~Adams and E.~I.~Blount, J.~Chem.~Phys.~{\bf 10}, 286 (1959).
\bibitem{Blount}
E.\ I.\ Blount, in {\it Solid State Physics} edted by
F.~Seitz and D.~Turnbull (Academic Press, NY, 1962) {\bf 13}, p.~305.
\bibitem{Chang1}
M.~C.~Chang and Q.~Niu, \prl{75}{1348}{95}.
\bibitem{Chang2}
M.~C.~Chang and Q.~Niu, \prb{53}{7010}{96}.
\bibitem{Berry}
M.~V.~Berry, Proc.~R.~Soc.~London, Ser.~A {\bf 392}, 45 (1984).
\bibitem{Simon}
B.~Simon, \prl{51}{2167}{83}.
\bibitem{Zak}
J.~Zak, \prb{40}{3156}{89}.
\bibitem{eff-ham}
W.~Kohn, \pr{115}{1460}{59}; E.~I.~Blount, {\it ibid} {\bf 126},
1636 (1962); L.~M.~Roth, J.~Chem.~Phys.~{\bf 23}, 433 (1962).
Wannier and Fredkin also developed an effective Hamiltonian
formalism; however, their method does not give corrections
explicitly as a power series in field strengths;
G.~H.~Wannier, \pr{117}{432}{60}; G.~H.~Wannier and D.~R.~Fredkin,
{\it ibid} {\bf 125}, 1910 (1962); G.~H.~Wannier, \rmp{34}{645}{62}.
\bibitem{Brown-67}
E.~Brown, \pr{166}{626}{67}.
\bibitem{Z-kq-rep}
J.~Zak, \prb{15}{771}{77}; \prb{16}{4154}{77}.
\bibitem{Bellissard}
R.~Rammal and J.~Bellissard, J.~Phys.~France {\bf 51}, 1803 (1990).
\bibitem{Wil-Heff}
M.~Wilkinson and R.~J.~Kay, \prl{76}{1896}{96}.
\bibitem{Harper}
P.~G.~Harper, Proc.~R.~Soc.~London, Ser.~A{\bf 86}, 874 (1955).
\bibitem{incom-S85}
J.~B.~Sokoloff, Phys.~Rep.~{\bf 126}, 189 (1985).
\bibitem{Nenciu}
G.~Nenciu, \rmp{63}{91}{91}.
\bibitem{tdvp-K}
P.~Kramer and M.~Saraceno, {\it Lecture Notes in Physics} {\bf 140}:
{\it Geometry of the Time-Dependent Variational Principle in
Quantum Mechanics}, (Springer-Verlag, 1981).
\bibitem{argendu}
A.~K.~Pattanayak,  Ph.D.~Dissertation, The University
of Texas, Austin, 1994.
\bibitem{Gen-Wf}
See Ref.~\onlinecite{Nenciu}. However, the existence of a
generalized type of Wannier functions with the required
localization properties has been proved by M. Wilkinson, J.\ Phys.\ :
Condensed Matter  {\bf 10}, 7407 (1998).
\bibitem{Balki}
G.~Venkataraman, D.~Sahoo and V.~Balakrishnan,
{\it Beyond the Crystalline State - An Emerging Perspective}
(Springer-Verlag, 1989).
\bibitem{Geller}
M.~R.~Geller and W.~Kohn, Phys.~Rev.~Lett.~{\bf 70}, 3103 (1993).
\bibitem{condition}
The time scale associated with an energy gap is given by the
uncertainty time $h/E_g$, where $E_g$ is the gap size. The length
scale of an gap is given by $1/\kappa_{\rm m}$, where
$\kappa_{\rm m}$ is the maximum size of the imaginary part of the
wave vector in the gap.
\bibitem{WPD}
Our  results are valid  only in the time scale in  which
the effects of spreading may be ignored.
For studies on the dynamics of higher order
moments of  a wave packet, see Ref.~\onlinecite{argendu}. A
systematic study of wave packet dynamics was initiated by
  E.~J.~Heller, J.~Chem.~Phys.~{\bf 62}, 1544 (1975).
\bibitem{symmetrzn}
This result can be verified for Hamiltonians of the form
\( \Ham = ({1}/{2m})(\hbp+e\bA(\hbx,t))^2 +
V[\hbx,\{\beta_i(\hbx,t)\}]\), the only types considered in this
paper.
\bibitem{Dirac}
P.~A.~M.~Dirac, {\it Lectures on Quantum Mechanics},
Belfer Graduate School of Science, Yeshiva University,
New York, (1964), and Can.~J.~Math.~{\bf 2}, 129 (1950).
\bibitem{tabor}
Michael Tabor, {\it Chaos and Integrability in Nonlinear
Dynamical Systems: An Introduction}
(Wiley Interscience Publication,  New York, 1989).
\bibitem{Wil-Bp}
M.~Wilkinson, J.~Phys.~A: Math.~Gen.~{\bf 17}, 3459 (1984).
\bibitem{pathIQR}
H.~Kuratsuji and S.~Iida, Prog.~Theor.~Phys.~{\bf 74}, 439  (1985).
%\bibitem{gutzwiller}
%M.~Gutzwiller, {\it Chaos in Classical and Quantum Mechanics},
%Springer-Verlag (New York) 1990.
\bibitem{mag.en}
In addition to the works cited in Ref.~\onlinecite{eff-ham} see
L.~M.~Roth, \pr{118}{1534}{60}; E.~H.~Cohen and E.~I.~Blount,
Phil.~Mag.~{\bf 5}, 115 (1960),
derive the orbital magnetic moment
for an effective mass Hamiltonian.
\bibitem{tktn}
D.~J.~Thouless, M.~Kohmoto, M.~P.~Nightingale
and M.~den Nijs, Phys.~Rev.~Lett.~{\bf 49}, 405 (1982).
Q.~Niu, D.~J.~Thouless and Y.~Wu, Phys.~Rev.~B {\bf 31}, 3372 (1985).
\bibitem{zak-ph}
J.~Zak, Phys.~Rev.~Lett.~{\bf 62}, 2747 (1989) and Ref.~\onlinecite{Zak}.
For centro-symmetric crystals, the
Berry phase is quantized and equals $0$ or $\pi$;
this  corresponds to centers of Wannier functions being either at
$x= la \mbox{ or } (l+1/2)a$, where $l$ is an integer.
\bibitem{WSL}
G.~H.~Wannier, \pr{117}{432}{60}. Though Wannier-Stark ladder was
predicted in the early 1960's,
only in the late 1980's that it was confirmed by experiments.
For details, see
E.~E.~Mendes and G.~Bastard, Phys.~Today {\bf 46} (6), 34 (1993).
\bibitem{zak-wsl-Bp}
J.~Zak, \prl{20}{1477}{69}.
\bibitem{lax}
M.~Lax, (unpublished).
\bibitem{abs-dp}
L.~Kleinman, \prb{24}{7412}{81}, has pointed out the difficulty in
defining absolute potentials and deformation potentials for infinite
crystals.
We defer a more detailed  consideration to a future publication.
\bibitem{df-el}
This definition  corresponds to  the
Eulerian convention.  In the Lagrangian convention one
defines another  displacement field  $\bu^{\SS{L}}(\bx)$
such that $\bu^{\SS{L}}(\bR_l) = {\bf u}_l$.
In either convention,  strains are of the same
order of magnitude which makes comparison with
other works reasonable.
\bibitem{D-P}
J.~Bardeen and W.~Shockley, \pr{80}{72}{50}, introduced the concept of
deformation potential as the strain  induced shift in electron energy  to
simplify the computation of the matrix elements of the
electron-phonon interaction at long wavelengths.
%\bibitem{scale-T}
%L.~J.~Sham and J.~M.~Ziman, Solid State Physics, {\bf 15}, 221 (1963).
\bibitem{DPT-KA}
F.~S.~Khan and P.~B.~Allen, \prb{29}{3341}{84}, use
Fermi energy as the reference point rather than the
band bottom in their definition of the deformation
potential, and include a term for the shift in the
Fermi energy due the applied strain.
However, this reference can be incorporated
at the level of Hamiltonian itself by a suitable mapping,
which result in band energies modified accordingly.
\bibitem{D-P-T}
Referred to as the deformation potential theorem,
the result was first derived by Bardeen and Shockley
for Bloch states near band
edges where the term
$m {\bf v} {\bf v}$ is zero (Ref.~\onlinecite{D-P}).
The exact result (\ref{dpt-wb})
was obtained by Khan and Allen(Ref.~\onlinecite{DPT-KA}) and later
confirmed by E.~Kartheuser and S.~Rodriguez, \prb{33}{772}{86}.
\bibitem{niu-ct}
Q.~Niu, \prl{64}{1812}{90}, and Mod.~Phys.~Lett.~B {\bf 5}, 923
(1991).
\bibitem{bc-zero}
This can always be done as long as the Berry curvature
$\tensor{{\Omega}}_{\bk \bk}$ is zero everywhere.
This condition is satisfied for centro-symmetric crystals in the
absence of magnetic fields.
\bibitem{A-B-effect}
Y.~Aharonov and D.~Bohm, \pr{115}{485}{59}.
\bibitem{A-B-scatt}
K.~Kawamura,  Z.~Physik {\bf B 29}, 101 (1978).
\bibitem{Bp-Sdisl}
D.~M.~Bird and A.~R.~Preston, \prl{61}{2863}{88}.
See Eqs.~4 (a), and 4 (b) for the calculated Berry phase.
\bibitem{KSP-Resta}
R.~Resta and  L.~Colombo, \prb{41}{12358}{90}.
\end{references}
\end{document}